\documentclass[preprint,showpacs,preprintnumbers,amsmath,amssymb]{revtex4}
\usepackage[abs]{overpic}
\usepackage{graphicx,subfigure}
\begin{document}
\preprint{HUPD1101}
\def\tbr{\textcolor{red}}
\def\tcr{\textcolor{red}}
\def\ov{\overline}
\def\dprime{{\prime \prime}}
\def\nn{\nonumber}
\def\f{\frac}
\def\beq{\begin{equation}}
\def\eeq{\end{equation}}
\def\bea{\begin{eqnarray}}
\def\eea{\end{eqnarray}}
\def\bsub{\begin{subequations}}
\def\esub{\end{subequations}}
\def\dc{\stackrel{\leftrightarrow}{\partial}}
\def\ynu{y_{\nu}}
\def\ydu{y_{\triangle}}
\def\ynut{{y_{\nu}}^T}
\def\ynuv{y_{\nu}\frac{v}{\sqrt{2}}}
\def\ynuvt{{\ynut}\frac{v}{\sqrt{2}}}
\def\d{\partial}
\title{Quantum correction to tiny vacuum expectation value
in two Higgs doublet model for Dirac neutrino mass}
\author{
Takuya  Morozumi$^{(a)}$,
Hiroyuki Takata$^{(b)}$, and   
Kotaro Tamai$^{(a)}$ \\
}
\address{(a)
Graduate School of Science, Hiroshima University,
Higashi-Hiroshima, 739-8526, Japan \\
(b) Tomsk State Pedagogical University,
Tomsk, 634041, Russia \\}
\date{\today}
\def\nn{\nonumber}
\def\beq{\begin{equation}}
\def\eeq{\end{equation}}
\def\bei{\begin{itemize}}
\def\eei{\end{itemize}}
\def\bea{\begin{eqnarray}}
\def\eea{\end{eqnarray}}
\def\ynu{y_{\nu}}
\def\ydu{y_{\triangle}}
\def\ynut{{y_{\nu}}^T}
\def\ynuv{y_{\nu}\frac{v}{\sqrt{2}}}
\def\ynuvt{{\ynut}\frac{v}{\sqrt{2}}}
\def\s{\partial \hspace{-.47em}/}
\def\ad{\overleftrightarrow{\partial}}
\def\ss{s \hspace{-.47em}/}
\def\pp{p \hspace{-.47em}/}
\def\bos{\boldsymbol}
\begin{abstract}
We study a Dirac neutrino mass model of Davidson and Logan. 
In the model, the smallness
of the neutrino mass is originated from the small vacuum
expectation value of the second Higgs of two Higgs doublets. 
We study the one loop effective
potential of the Higgs sector and examine how the small vacuum 
expectation is stable under the radiative correction.
By deriving formulae of the radiative correction,
we numerically study how large the one loop correction is
and show how it depends on
the quadratic mass terms 
and quartic couplings of the Higgs potential. 
The correction changes depending on the various
scenarios for extra Higgs mass spectrum.
\end{abstract}
\maketitle
\section{Introduction}
The smallness of the neutrino
mass compared with the other quarks and leptons
are one of the mysteries of nature.
Recently, a new mechanism generating small Dirac mass terms
for neutrino
has been proposed \cite{Davidson:2009ha,Davidson:2010sf,Gabriel:2006ns}.
The similar mechanism  generating the small neutrino
Dirac mass term for the TeV seesaw mechanism is also proposed in
\cite{Ma:2000cc} and phenomenology is studied in
\cite{Haba:2010zi} and \cite{Haba:2011nb}. There are also models with radiatively generated Dirac mass term in 
\cite{Kanemura:2011jj, Nasri:2001ax}.
The interesting feature of the model proposed in \cite{Davidson:2009ha,Davidson:2010sf} is 
the tiny vacuum expectation
value for an extra Higgs SU(2) doublet \cite{Hashimoto:2004kz}.
The small neutrino
mass is realized without introducing tiny Yukawa coupling
for neutrinos.
A softly broken global U(1) symmetry guarantees the tiny 
vacuum expectation value for the extra doublet.
In addition to the small softly breaking mass parameter, 
the mass squared parameter for the
extra Higgs is chosen to be positive
so that the light pseudo Nambu Goldstone bosons
due to the softly broken global symmetry do not appear.
This is a contrast to
the mass squared parameter for the standard model like Higgs boson.

In the present paper, we study the global minimum 
of the tree level Higgs potential by explicitly solving the
stationary conditions.
There are many studies of the tree level Higgs potential of general 
two Higgs doublet model \cite{Ferreira:2004yd,Maniatis:2006fs,Barroso:2007rr,Ivanov:2007de,Ginzburg:2007jn,Ivanov:2008er}. (See
also \cite{Branco:2011iw} for recent review of two Higgs doublet model.) 
It has been shown that
the charge neutral vacuum is lower than the charge breaking vacuum
\cite{Ferreira:2004yd}.
Also the vacuum energy 
difference of two neutral minima was derived \cite{Barroso:2007rr,Ginzburg:2007jn}. We make use of the results and identify the vacuum of 
the present model. When the U(1) symmetry breaking term is turned off,
the tree level Higgs potential and the phase structure of the present model 
is rather similar to the model with $Z_2$ discrete symmetry 
\cite{Gustafsson:2011wg, Krawczyk:2011jy}.
In contrast to $Z_2$ symmetric case, it is essential 
to keep the soft breaking term when finding the true vacuum.
If we set the symmetry breaking term at zero, then the order parameter
corresponding to the softly broken U(1) symmetry becomes redundant parameter
and can not be determined. We treat the soft breaking term as small
expansion parameter and
obtain the vacuum expectation values
and the vacuum energies in terms of the parameters of the 
Higgs potential.

The constraints on the parameters of the model for which the desired
vacuum can be realized, are derived and they are rewritten  
in terms of Higgs masses and a few coupling constants which 
can not be directly related to the Higgs masses. 
These constraints are fully used when we study the radiative
corrections to the vacuum expectation values numerically. 

Beyond the tree level,
we study the radiative correction to the Higgs potential and
the vacuum expectation values 
of Higgs. 
Since the neutrino masses are proportional to
the vacuum expectation value of one of Higgs
, one can also compute the radiative
corrections to neutrino masses.
As already noted in \cite{Davidson:2009ha}
the radiative correction to the 
softly breaking mass parameter is logarithmically divergent
and it is renormalized multiplicatively. 
We derive the formulae 
for the one loop corrected vacuum expectation values
for two Higgs doublets by studying 
one loop corrected effective potential.
The corrections are evaluated numerically
by exploring the parameter regions allowed  
from the global minimum condition for the vacuum.
We show how the radiative corrections change depending on
the extra Higgs spectrum.
The radiative corrections are also evaluated for the case that 
a relation among the coupling constants is satisfied.

The paper is organized as follows. 
In section (II), we derive the condition for the 
desired vacuum being global minimum. In section (III) 
one loop effective potential is derived and
one loop corrections to the vacuum expectation values
are obtained in section (IV). 
In section (V), the corrections are evaluated numerically
for various choices of parameters of the Higgs potential.
Section (VI) is devoted to summary and discussion.

\section{Model for Dirac neutrino with a tiny vacuum expectation value}
The model of the Dirac neutrino is proposed in \cite{Davidson:2009ha}. 
In \cite{Davidson:2009ha}, two Higgs SU(2) doublets are introduced,
\bea
\Phi_1=\frac{1}{\sqrt{2}}\begin{pmatrix}
\phi_1^1+ i \phi_1^2 \\
\phi_1^3+ i \phi_1^4  
\end{pmatrix}, \quad 
\Phi_2=\frac{1}{\sqrt{2}} \begin{pmatrix} 
\phi_2^1+ i \phi_2^2 \\
\phi_2^3+ i \phi_2^4 \end{pmatrix},
\eea
where $\Phi_1$'s vacuum expectation value is nearly 
equal to the electroweak breaking scale and
the second Higgs $\Phi_2$ has a small vacuum expectation value
which gives rise to neutrino mass.
The Higgs potential in \cite{Davidson:2009ha} is,
\bea
V_{\rm tree}=\sum_{i=1,2}\left( m_{ii}^2 \Phi_i^\dagger \Phi_i+
\frac{\lambda_i}{2} (\Phi_i^\dagger \Phi_i)^2 \right)
-(m_{12}^2 \Phi_1^\dagger \Phi_2 + h.c.)+
\lambda_3  (\Phi_1^\dagger \Phi_1)(\Phi_2^\dagger \Phi_2)
+\lambda_4 |\Phi_1^\dagger \Phi_2|^2.
\label{eq:tree}
\eea
U(1)$^\prime$ charge is assigned to the second Higgs.  
The U(1)$^\prime$ global 
symmetry is broken softly with the term $m_{12}^2$.
In this paper, we introduce the following real O(4) representation
for each doublet,
because this parametrization is convenient when computing the
one loop corrected effective potential.
\bea
\phi_1^a=\begin{pmatrix}
\phi_1^1 \\
\phi_1^2 \\
\phi_1^3 \\
\phi_1^4 
\end{pmatrix}, \quad
\phi_2^a=
\begin{pmatrix}
\phi_2^1 \\
\phi_2^2 \\
\phi_2^3 \\
\phi_2^4 
\end{pmatrix}, \quad
\tilde{\phi}_1^a= 
\begin{pmatrix} 
-\phi_1^2 \\
\phi_1^1 \\
-\phi_1^4 \\
\phi_1^3 
\end{pmatrix}.
\label{eq:phi}
\eea
Using the notation above, the
tree level effective potential introduced in Eq.(\ref{eq:tree}) 
can be written as,
\bea
V_{\rm tree}&=&m_{11}^2 \frac{1}{2} \sum_{a=1}^4 (\phi_1^a)^2
+m_{22}^2 \frac{1}{2}  \sum_{a=1}^4 (\phi_2^a)^2
-m_{12}^2 \sum_{a=1}^4 \phi_1^a \phi_2^a \nn \\
&+& \frac{\lambda_1}{8} (\sum_{a=1}^4 {\phi_1^a}^2)^2
+ \frac{\lambda_2}{8} (\sum_{a=1}^4 {\phi_2^a}^2)^2 
+ \frac{\lambda_3}{4} (\sum_{a=1}^4 {\phi_1^a}^2)   
(\sum_{a=1}^4 {\phi_2^a}^2) \nn \\
&+& \frac{\lambda_4}{4} \left( (\sum_{a=1}^4 
\phi_1^a \phi_2^a)^2+ (\sum_{a=1}^4
\tilde{\phi}_1^a \phi_2^a)^2 \right),
\eea
where one can choose $m_{12}^2$ real and positive.
With the notation of Eq.(\ref{eq:phi}), the softly broken
global symmetry U(1)$^\prime$ corresponds to 
the following transformation on $\phi_2^a$,
\bea
\phi_2^\prime =O_{U(1)^\prime} \phi_2 &=&
\left(
\begin{array}{cccc}
\cos \phi & -\sin \phi & 0 & 0 \\
\sin \phi & \cos \phi & 0 & 0 \\
0 & 0 & \cos \phi & -\sin \phi \\
0 & 0 & \sin \phi & \cos \phi 
\end{array} \right) \phi_2.
\label{eq:U(1)}
\eea
$\phi_1$ does not transform under U(1)$^\prime$.
Therefore U(1)$^\prime$ is broken softly when $m_{12}^2$ does not vanish.
Without loss of generality, one can choose the vacuum expectation
values of Higgs with the form given as,
\bea
\langle \phi_1 \rangle = 
\begin{pmatrix} 0 \\
0\\
v \cos \beta \\
0 \\
\end{pmatrix},  \quad
\langle \phi_2 \rangle =
\begin{pmatrix} v \sin \beta \sin \alpha \cos \theta^\prime \\
                -v \sin \beta \sin \alpha \sin \theta^\prime \\
                v \sin \beta \cos \alpha \cos \theta^\prime \\
                -v \sin \beta \cos \alpha \sin \theta^\prime \end{pmatrix},
\label{eq:v2gene}
\eea
where the range for $\theta^\prime$ is $[0, 2 \pi)$ and 
the range  for $\beta$ and $\alpha$ is $[0, \frac{\pi}{2}]$.
We call the four order parameters as 
${\varphi}_I=(v,\beta, \alpha, \theta^\prime)$, $(I=1,2,3,4)$.
When $m_{12}$ vanishes, by taking $\phi=\theta^\prime$ in Eq.(\ref{eq:U(1)}), 
one can rotate
$\theta^\prime$ away in Eq.(\ref{eq:v2gene}). For the most general
case, in total, there are four independent order parameters
when U(1)$^\prime$ symmetry is broken.

For completeness of our discussion, we give the constraints
on the quartic couplings from condition that
the tree level potential is the bounded below, \cite{Davidson:2009ha}, \cite{Ferreira:2004yd}, \cite{Hill:1985tg}.
\bea
\lambda_1 > 0, \quad \lambda_2 >0 ,\label{eq:gcon0} \\
-\sqrt{\lambda_1 \lambda_2} \le \lambda_3, \label{eq:gcon1} \\
-\sqrt{{\lambda_1}{\lambda_2}} \le \lambda_3+\lambda_4. \label{eq:gcon2}
\eea
In addition to the conditions on the quartic terms,
one can constrain the parameters including the quadratic terms
so that the desired vacuum satisfies the global minimum conditions
of the potential.  About the global minimum of the tree potential, 
it was shown that the energy of charge neutral vacuum is lower than that of the charge breaking vacuum \cite{Ferreira:2004yd}.
We therefore set $\alpha$ zero. 
We also require the vacuum expectation value
of the second Higgs is much smaller than that of the first Higgs,
which implies that $\tan \beta$ is small.  
In terms of the parametrization in Eq.(\ref{eq:v2gene}) with $\alpha = 0$, 
the potential can be written as,
\bea
V_{\rm tree}(v,\beta, \theta^\prime)
= A(\beta) v^4+ B(\beta, \theta^{\prime}) v^2,
\eea
where,
\bea
A(\beta) &=& 
\frac{ \lambda _1 }{8} \cos ^4 \beta + \frac{ \lambda _2}{8} \sin ^4 \beta 
+\left( \frac{\lambda _3}{4} + \frac{\lambda _4}{4} \right) \cos ^2 \beta \sin ^2 \beta, \nn \\
B(\beta, \theta^\prime) &=& \frac{m^2_{11}}{2} \cos ^2 \beta + \frac{m^2_{22}}{2} \sin ^2 \beta 
-m^2_{12} \cos \theta ^\prime \cos \beta \sin \beta.
\eea
We first find the global minimum of $V_{\rm tree}$. 
The stationary conditions $
\frac{\partial V_{\rm tree}}{\partial \varphi_I}=0$ $(I=1,2,4)$,
are written as,
\bea
&& v(2 A v^2+B)=0, \label{eq:v2} \\
&& 2 r_4=\sin 2 \beta \frac{(1-r_1 r_2) \cos 2 \beta +r_2-r_1 r_3}
{r_2 \cos^2 2 \beta +(r_3+1) \cos 2 \beta +r_2}, \label{eq:beta}\\
&& m_{12}^2\sin \theta^\prime \sin 2\beta = 0, 
\label{eq:thetap} 
\eea
where $r_i(i=1 \sim 4)$ are defined as,
\bea
r_1 &=& \frac{m^2_{11}-m^2_{22}}{m^2_{11}+m^2_{22}}, \nn \\
r_2 &=& \frac{\lambda _1 -\lambda _2}{\lambda _1+\lambda _2-2\lambda _3-2\lambda _4}, \nn \\
r_3 &=& \frac{\lambda _1+\lambda _2+2\lambda _3+2\lambda _4 }{\lambda _1+\lambda _2-2\lambda _3-2\lambda _4}, \nn \\
r_4 &=& \frac{m^2_{12} \cos \theta^\prime }{m^2_{11}+m^2_{22}}.
\eea
The stationary conditions Eq.(\ref{eq:v2}) and Eq.(\ref{eq:beta}) correspond 
to Eq.(36) of \cite{Ginzburg:2007jn}.
Here we solve them explicitly by treating the soft breaking term 
$m_{12}$ as perturbation. 
The non-zero solution for $v^2$
in  Eq.(\ref{eq:v2}) is written as,
\bea
v^2&=&-\frac{B}{2A}=-4\frac{m_{11}^2+m_{22}^2}{\lambda_1+\lambda_2
-2 \lambda_{34}}\frac{1+r_1 \cos 2 \beta-2 r_4 \sin 2 \beta}
{\cos^2 2 \beta+r_3+2 r_2 \cos 2 \beta},
\eea
where $\lambda_{34}=\lambda_3+\lambda_4$.
Substituting  it into
$V_{\rm tree}$, one obtains, 
\bea
V_{\rm tree} &\ge& V_{\rm  min.}=-\frac{(m_{11}^2+m_{22}^2)^2}{2(\lambda_1+\lambda_2-2\lambda_{34})} \frac{(1+r_1 \cos 2 \beta-2r_4 \sin 2 \beta)^2}{\cos^2 2 \beta+2 r_2
\cos 2 \beta +r_3}.
\eea 
For non-zero 
$m_{12}^2$ and $\sin 2 \beta$, the solution of Eq.(\ref{eq:thetap}) is 
$\sin \theta^\prime=0$.
One still needs to find $\beta$
among the solutions of Eq.(\ref{eq:beta}),
which leads to the minimum of $V_{\rm min.}$.
We solve Eq.(\ref{eq:beta}) and determine $\beta$
by treating $r_4 \ (m_{12}^2)$
as a small expansion parameter.
One can easily find the approximate solutions as,
\bea
\begin{cases}
(1) \sin \beta= \frac{\lambda_1 m_{12}^2}
{|m_{22}^2 \lambda_1-m_{11}^2 \lambda_{34}|}, \quad 
\cos \theta^\prime= \rm{sign}(m_{22}^2 \lambda_1-m_{11}^2 \lambda_{34}),
\\
(2) \cos \beta= \frac{\lambda_2 m_{12}^2}
{|m_{11}^2 \lambda_2-m_{22}^2 \lambda_{34}|}, 
\quad 
\cos \theta^\prime= \rm{sign}(m_{11}^2 \lambda_2-m_{22}^2 \lambda_{34}), 
\\
(3) \cos 2 \beta=\frac{m_{11}^2 (\lambda_{34}+ \lambda_2)-m_{22}^2
(\lambda_{34}+\lambda_1)}{m_{11}^2 (-\lambda_{34}+ \lambda_2)+m_{22}^2
(-\lambda_{34}+\lambda_1)}+O(r_4),
\end{cases}
\label{eq:solution}
\eea

Corresponding to each solution, (1)$\sim$(3) of Eq.(\ref{eq:solution}), 
the vacuum expectation value $v^2$ and 
the minimum of the potential are obtained.
\bea
&&(v^2,V_{\rm min})= \nn \\
&&\begin{cases}
(1) \left(-\frac{2m_{11}^2}{\lambda_1}+ 2 \lambda_1(m_{22}^2-m_{11}^2)
\left(\frac{m_{12}^2}
{m_{22}^2 \lambda_1-m_{11}^2 \lambda_{34}}\right)^2,
 -\frac{m_{11}^4}{2 \lambda_1} + 
 \frac{m_{12}^4 m_{11}^2}{m_{22}^2 \lambda_1-
m_{11}^2 \lambda_{34}}\right), \nn \\
(2) \left(-\frac{2m_{22}^2}{\lambda_2}+ 2 \lambda_2(m_{11}^2-m_{22}^2)
\left(\frac{m_{12}^2}
{m_{11}^2 \lambda_2-m_{22}^2 \lambda_{34}}\right)^2,
-\frac{m_{22}^4}{2 \lambda_2} +
 \frac{m_{12}^4 m_{22}^2}{m_{11}^2 \lambda_2-
m_{22}^2 \lambda_{34}}\right),\nn \\
(3)\left(2\frac{(\lambda_{34}-\lambda_2)m_{11}^2+
(\lambda_{34}-\lambda_1)m_{22}^2}
{\lambda_1 \lambda_2-\lambda_{34}^2}+O(r_4),
-\frac{\lambda_2 m_{11}^4-2 m_{11}^2 m_{22}^2 \lambda_{34} +
 \lambda_1 m_{22}^4}
{2(\lambda_1 \lambda_2-\lambda_{34}^2)}+O(r_4)\right). 
\end{cases} \nn \\
\label{eq:v2Vmin}
\eea
The leading terms of the vacuum expectation values 
agree with those obtained in $Z_2$ symmetric model \cite{Krawczyk:2011jy}.
If $\sin 2 \beta=0$, then $r_4$ must be vanishing and 
$\cos\theta^\prime=0$ from Eq.(\ref{eq:beta})
and Eq.(\ref{eq:thetap}).
\begin{table}
\begin{tabular}{|c|c|} \hline
(1) $\sin \beta=$O$(r_4)$   & $-\frac{m_{11}^4}{2 \lambda_1}-
\frac{m_{12}^4}{\lambda_3+\lambda_4-\frac{m_{22}^2}{m_{11}^2} \lambda_1}$   
\\ \hline
(2) $\cos \beta=$O$(r_4)$ & $-\frac{m_{22}^4}{2 \lambda_2}-
\frac{m_{12}^4}{\lambda_3+\lambda_4-
\frac{m_{11}^2}{m_{22}^2} \lambda_2}$ 
\\ \hline 
(3)$\cos 2 \beta=$O$(1)$&$-
\frac{\lambda_1 m_{11}^4 -2 m_{11}^2 m_{22}^2 (\lambda_3+\lambda_4)
+\lambda_2 m_{22}^4}{2(\lambda_1 \lambda_2-(\lambda_3+\lambda_4)^2)}$
\\ \hline
\end{tabular}
\caption{Classification of the solutions with non zero $\sin 2 \beta$
of the stationary conditions of Higgs potential. For (3), O($r_4$) correction
is not shown.}
\label{table:nonzerobeta}
\end{table}
\begin{table}
\begin{tabular}{|c|c|} \hline
   &$\cos \theta^\prime=0$  \\ \hline
(4)$\sin \beta=0$&$-\frac{m_{11}^4}{2 \lambda_1}$
\\ \hline
(5)$\cos \beta=0$&$-\frac{m_{22}^4}{2 \lambda_2}$ \\ \hline 
\end{tabular}
\caption{Classification of the solutions with $\sin 2 \beta=0$.}
\label{table:betazero}
\end{table}
The vacuum energies of the 
non-zero $\sin 2 \beta$ solutions are shown in  
Tables \ref{table:nonzerobeta}.
In Table \ref{table:betazero}, the vacuum energies of 
the solutions with $\sin 2 \beta=0$ are summarized.

Next we derive the constraints on the parameters so that the solution corresponding to (1) in Table \ref{table:nonzerobeta} becomes the global minimum 
of the potential.
Since the other cases (2)-(5) do not have desired properties
, we restrict the parameter space so that these solutions can not be a global minimum. Since $v$ must have large positive
vacuum expectation
value, $m_{11}^2$ must be negative.
In order that the vacuum energy of (1) is lower than that of (4)
, 
\bea
m_{22}^2 \lambda_1-m_{11}^2 \lambda_{34} >0 ,\ (\cos \theta^\prime=1).
\label{eq:thetap0}
\eea
When Eq.(\ref{eq:thetap0}) is satisfied and the solution (1) does exist,
one can show that the vacuum energy of solution (3) is higher than that of
(1). 
Furthermore when $m_{22}^2 > 0$, the solutions corresponding to 
(2) and (5) are not realized.
Then  one can state the region of parameter space which is consistent 
with the case that the vacuum 
(1) becomes global minimum is, 
\bea
m_{11}^2 <0 ,\ m_{22}^2 >0,\ \lambda_{34}>
\frac{m_{22}^2}{m_{11}^2} \lambda_1. \ 
\label{eq:lcon1}
\eea
Next we consider the case with negative $m_{22}^2$.
In this case we impose the additional condition so that the vacuum energies
corresponding to (2) and (5) are higher than that of (1).
\bea
\frac{m_{11}^4}{\lambda_1}> \frac{m_{22}^4}{\lambda_2}.
\label{eq:c12}
\eea
Then the condition for (1) is global minimum in this case is,
\bea
m_{11}^2 <0 ,\ m_{22}^2 <0,\lambda_{34}>
\frac{m_{22}^2}{m_{11}^2} \lambda_1,\ 
\lambda_2 \frac{m_{11}^2}{m_{22}^2}
>\lambda_1 \frac{m_{22}^2}{m_{11}^2}. 
\label{eq:lcon3}
\eea
In the following sections, we explore the regions for the parameters
obtained in Eq.(\ref{eq:lcon1}), Eq.(\ref{eq:lcon3}),
Eq.(\ref{eq:gcon1}) and Eq.(\ref{eq:gcon2}).
\section{Effective potential in one loop and Renormalization}
In this section, 
we derive the effective potential within one loop
approximation.
We introduce a real scalar fields with eight components as,
$
\phi^i=(\phi_1^1, \phi_1^2, \phi_1^3, \phi_1^4,
\phi_2^1, \phi_2^2,\phi_2^3, \phi_2^4)^T$,$(i=1 \sim 8)$.
With the notation above, the one loop effective action is given
as,
\bea
\Gamma^{1 {\rm loop}}_{\rm eff}&=&
i\frac{1}{2} \ln {\rm det}
D^{-1}(\phi),\nn \\
D^{-1}&=&\Box+M_T^2,
\eea
where $M_T^2$ is the mass squared matrix of the Higgs potential, 
\bea
M_T^2&=&M^2(\phi)+\begin{pmatrix} m_{11}^2\times 1 & 0 \\
0 & m_{22}^2 \times 1 \end{pmatrix}-m_{12}^2 \sigma_1, \nn \\
M^2(\phi)_{ij}&=&
\frac{\partial^2 V_{\rm tree}^{(4)}}{\partial \phi_i \partial \phi_j},
\label{eq:MTMphi}
\eea
where $1$($0$) denotes $4 \times 4$ unit (zero) matrix.
$\sigma_1$ is defined as,
\bea
\sigma_1=
\begin{pmatrix}
0 & 1 \\
1 & 0 \end{pmatrix}. 
\label{eq:sigma}
\eea
In Eq.(\ref{eq:sigma}),
$1$($0$) also denotes a four by four unit(zero) matrix. 
In modified minimal subtraction scheme, the finite part of the one loop effective potential becomes, 
\bea
V_{\rm 1 loop}&=&\frac{\mu^{4-d}}{2}
\int \frac{d^d k}{(2 \pi)^d i} {\rm TrLn}(M_T^2-k^2)+V_c,  \nn \\
&=&\frac{1}{64 \pi^2} {\rm Tr}.\left(M_T^4 
({\rm Ln} \frac{M_T^2}{\mu^2}-\frac{3}{2})\right).
\label{eq:V1loop}
\eea
$V_c$ denotes the counterterms and the derivation of $V_c$ can be found in Appendix A.
\section{One loop corrections to the vacuum expectation values}
In this section, we compute the one loop corrections to the
vacuum expectation values. Using the symmetry
of the model, in general, one can choose ${\varphi}_I=
(v,\beta, \alpha, \theta^\prime)$ as the vacuum expectation values
of Higgs potential.   Their values
are obtained as the stationary points 
of the one loop corrected effective potential 
$V=V_{\rm tree}+V_{1 {\rm loop}}$,
\bea
\frac{\partial V}{\partial {\varphi}_I}=0.
\eea
By denoting the vacuum expectation values as sum of the tree level ones and the
one loop corrections to them;
$\varphi_I=\varphi^{(0)}_I+\varphi^{(1)}_I$, one obtains the 1 loop 
corrections,
\bea
\varphi_I^{(1)}&=&-(L^{-1})_{IJ}
\frac{\partial V_{\rm 1 loop}}{\partial \varphi_J}
\Biggl{|}_{\varphi=\varphi^{(0)}},\nn \\
&=&-\frac{1}{32 \pi^2}(L^{-1})_{IJ} \sum_{i=1}^8
(O^T 
\frac{\partial M^2}{\partial \varphi_J}\Biggl{|}_{\varphi=\varphi^{(0)}} 
\hspace{-8mm}O)_{ii} {M_D}_i^2 (\ln \frac{{M_D}_i^2}{\mu^2}-1),
\label{eq:ordersimp}
\eea
where ${M_D}^2$ is a diagonal $ 8 \times 8$ tree level mass squared matrix of Higgs sector and $L_{IJ}$ is $ 4 \times 4$ matrix 
given by the second derivatives of the tree level Higgs potential
with respect to the order parameters, 
\bea
L_{IJ}&=&\frac{\partial^2 V_{\rm tree}}{\partial 
\varphi_I \partial \varphi_J}\Biggl{|}_{\varphi=\varphi^{(0)}}.
\label{eq:LIJ} 
\eea
The diagonal Higgs mass matrix squared $M_D^2$ is related to
$8 \times 8$ Higgs mass matrix squared $M_T^2$ in Eq.(\ref{eq:MTMphi}). 
\bea
O^T M_{T0}^2 O= M_D^2= \begin{pmatrix}
M_{H^+}^2 & 0 & 0 & 0 & 0 & 0 & 0 & 0\\
0  & M_{H^+}^2 & 0 & 0 & 0 & 0 & 0 & 0\\
0  & 0 & M_{A}^2 & 0 & 0 & 0 & 0 & 0 \\
0 & 0 & 0 & M_h^2 & 0 & 0 & 0 & 0 \\
0 & 0 & 0 & 0 & M_H^2 & 0 & 0 & 0 \\
0 & 0 & 0 & 0 & 0 & 0 & 0 & 0 \\
0 & 0 & 0 & 0 & 0 & 0 & 0 & 0 \\
0 & 0 & 0 & 0 & 0 & 0 & 0 & 0 \\
\end{pmatrix},
\label{eq:Higgsmass}
\eea
where $M_{T0}^2$ is obtained by substituting the vacuum expectation values to
$M_T^2$. $O$ is shown in appendix D. 
Since ${M_D}$ is the $8 \times 8$ diagonal 
matrix which elements correspond to
the Higgs masses and zero mass of the would be 
Nambu-Goldstone bosons, one may write Eq.(\ref{eq:ordersimp})
in a simple form.
The Higgs masses squared in Eq.(\ref{eq:Higgsmass}) are given by,
\bea
M_{H^+}^2&=&\frac{1}{2} \Biggl{[}\frac{1}{8} \left(\lambda _1+\lambda _2+6 \lambda _3-2 \lambda _4-\cos (4 \beta ) \left(\lambda _1+\lambda _2-2 \left(\lambda _3+\lambda _4\right)\right)\right) v^2 \nn \\
&+& 
(1-\cos (2 \beta )) m_{11}^2+(\cos (2 \beta )+1) m_{22}^2+2 \sin (2 \beta ) m_{12}^2
\Biggr{]}, \nn \\
M_A^2&=&M_{H^+}^2+\frac{\lambda_4 v^2}{2}, \nn \\
\frac{M_{h}^2+M_{H}^2}{2}&=&
\frac{1}{4} 
\left(\left(3 \lambda _1 \cos ^2(\beta )+
3 \sin ^2(\beta ) \lambda _2+\lambda _3+\lambda _4\right) v^2
+2 m_{11}^2+2 m_{22}^2\right), \nn \\
\frac{M_{H}^2-M_{h}^2}{2}&=&
\frac{1}{8} \Biggl{[}\{ 6 \cos (2 \gamma ) \left(\cos ^2(\beta ) \lambda _1-\sin ^2(\beta ) \lambda _2\right) \nn \\
&+&
(\cos (2 (\beta +\gamma ))-3 \cos (2 (\beta -\gamma ))) 
\left(\lambda _3+\lambda _4\right)\} v^2 \nn \\
&+& 4 \cos (2 \gamma ) 
m_{11}^2-4 \cos (2 \gamma ) m_{22}^2+8 \sin (2 \gamma ) m_{12}^2\Biggr{]},
\eea
where $\gamma$ is an angle with which one can 
diagonalize the $2 \times 2$ mass matrix for
CP even neutral Higgs.
$\tan 2 \gamma$ is given as,
\bea
\tan 2\gamma=
\frac{-4m_{12}^2+2\sin 2 \beta (\lambda_3+\lambda_4) v^2}
{(3(- \lambda_1 \cos^2 \beta + \lambda_2 \sin^2 \beta)
+\cos 2 \beta (\lambda_3+\lambda_4)) v^2-2(m_{11}^2-m_{22}^2)}.  
\eea
To compute Eq.(\ref{eq:ordersimp}), we still need to
calculate
$O^T \frac{\partial M^2}{\partial \varphi_I} O$ and $L_{IJ}$. They
are shown in appendix C.
Using Eq.(\ref{eq:ordersimp}) and Eq.(\ref{eq:althp}), 
one can find the quantum corrections for $\alpha$ and $\theta^\prime$
vanish,
\bea
\alpha^{(1)}=0, \  \theta^{\prime(1)}=0.
\eea
For $v^{(1)}$ and $\beta^{(1)}$, one obtains,
\bea
v^{(1)}&=&-\frac{1}{32 \pi^2}\frac{1}{\det L^\prime} \left(L_{22}\sum_{j=1}^5 
[O^T \frac{\partial M^2}{\partial \varphi_1} O]_{jj} M_{Dj}^2 
(\ln \frac{M_{Dj}^2}{\mu^2} -1)\right. \nn \\
&-& \left.     
L_{12}\sum_{j=1}^5 [O^T
\frac{\partial M^2}{\partial \varphi_2} O]_{jj} M_{Dj}^2 
(\ln \frac{M_{Dj}^2}{\mu^2} -1) \right), \nn \\
\beta^{(1)}&=&-\frac{1}{32 \pi^2} \frac{1}{\det L^\prime}
\left(-L_{12} \sum_{j=1}^5 [O^T \frac{\partial M^2}{\partial \varphi_1} O]_{jj} M_{Dj}^2 
(\ln \frac{M_{Dj}^2}{\mu^2} -1)\right. \nn \\
&+& \left. L_{11} \sum_{j=1}^5 
[O^T \frac{\partial M^2}{\partial \varphi_2} O]_{jj} M_{Dj}^2 
(\ln \frac{M_{Dj}^2}{\mu^2} -1) \right),
\label{eq:qc}
\eea
where $L^\prime$ is,
\bea
L^\prime=\begin{pmatrix} L_{11} & L_{12} \\
                  L_{12} & L_{22} \\
                  \end{pmatrix}.
\eea
The elements of $L^{\prime}$ are shown in Eq.(\ref{eq:elementofL}).
Eq.(\ref{eq:qc}) corresponds to the
 one loop exact formulae and is a main result
of present paper.
In the leading order of the expansion with  respect to
the symmetry breaking term $m_{12}^2$, the correction to
$v$ becomes,
\bea
v^{(1)}=&-&\frac{v}{32 \pi^2}
\Biggl\{3 \lambda_1 \left(\ln \frac{M_H^{2}}{\mu^2}-1 \right)
+2 \lambda_3 \frac{M_{H^{+}}^2}{M_H^2}
\left(\ln \frac{M_{H^{+}}^2}{\mu^2}-1\right) \nn \\
&+& (\lambda_3+\lambda_4) \left(\frac{M_A^{2}}{M_H^{2}}
\left(\ln \frac{M_A^{2}}{\mu^2}-1\right) +\frac{M_h^2}{M_H^2}
\left(\ln \frac{M_h^{2}}{\mu^2}-1\right) \right) \Biggr\}.
\label{eq:qcva}
\eea
The Higgs masses in the formulae are the ones in the 
limit of $m_{12} \rightarrow 0$,
\bea
M_{H}^2 &\simeq& m_{11}^2+\frac{3}{2} \lambda_1 v^2, \nn \\
M_A^2 \simeq M_{h}^2 &\simeq& m_{22}^2+\frac{\lambda_3+\lambda_4}{2}
v^2,\nn \\
M_{H^{+}}^2 &\simeq& m_{22}^2 +\frac{\lambda_3}{2} v^2,  
\label{eq:higgsmass}
\eea
where $v$ is related to $m_{11}^2$ as,
\bea
\frac{\lambda_1}{2} v^2 \simeq -m_{11}^2. 
\eea
The approximate formulae for the physical Higgs masses in 
Eq.(\ref{eq:higgsmass})
which are 
valid the limit
$m_{12} \rightarrow 0$, agree with the ones given in 
\cite{Davidson:2009ha} except the notational difference
of $M_H$ and $M_h$ \footnote{We denote $M_H$ as the
standard model like Higgs while in \cite{Davidson:2009ha},
it is called as $M_h$.}. 
The one loop correction to $\beta$ in the leading order
expansion of $m_{12}^2$ is given as,
\bea
&&\beta^{(1)}=-\frac{\beta}{32 \pi^2} \times \nn \\ 
&&\Biggr\{2\left(\lambda_2-\lambda_4-
\frac{\lambda_{3}(\lambda_3+\lambda_4)}{\lambda_1} \right)\frac{M_{H^+}^2}{M_A^2}
\left(\ln \frac{M_{H^+}^2}{\mu^2}-1 \right)
+(\lambda_2-\frac{(\lambda_3+\lambda_4)^2}{\lambda_1})\left(\ln \frac{M_{A}^2}{\mu^2}-1\right) \nn \\
&+&\left(3 \lambda_2 + (2 \Gamma-\frac{\lambda_3+\lambda_4}{\lambda_1})
(\lambda_3+\lambda_4)\right) 
\frac{M_h^2}{M_A^2} \left(\ln \frac{M_{h}^2}{\mu^2}-1\right) \nn \\
&-& 2(1+\Gamma)(\lambda_3+\lambda_4)
\frac{M_H^2}{M_A^2}\left( \ln \frac{M_H^2}{\mu^2}-1 \right)
\Biggr\},
\label{eq:qcba}
\eea
where,
\bea
\Gamma&=&\lim_{m_{12} \rightarrow 0} \frac{\gamma}{\beta}, \nn \\
      &=&\frac{M_A^2-M_H^2 \frac{\lambda_3+\lambda_4}{\lambda_1}}{M_H^2-M_A^2}.
\eea
Eq.(\ref{eq:qcba}) shows that the quantum correction is also proportional
to the soft breaking parameter $m_{12}^2$ which is expected.
We also note that the correction depends on the Higgs mass
spectrum
and quartic couplings.  The correlation to Higgs
spectrum is studied 
in the next section.
\section{Numerical Calculation}
In this section, we study the quantum correction to $\beta$ and $v$
numerically.
As shown in Eq.(\ref{eq:qcva}) and Eq.(\ref{eq:qcba}), the quantum corrections 
are written with  four Higgs masses  and the four quartic couplings. 
Since the neutral CP even and CP odd Higgs of the
second Higgs doublet are degenerate
as $M_A=M_h$ in the limit $m_{12} \rightarrow 0$ (See Eq.(\ref{eq:higgsmass})),
the three
Higgs masses $(M_H, M_A, M_{H^+})$ are independent. 
Moreover for a given charged Higgs mass and 
neutral Higgs mass,  $\lambda_1$ and $\lambda_4$ are given as,
\bea
\lambda_1&=&\frac{M_H^2}{v^2}, \nn \\
\lambda_4&=&2 \frac{M_A^2-M_{H^+}^2}{v^2}.
\eea
$\lambda_2$ and $\lambda_3$ are the remaining parameters to be fixed.
The lower limit of $\lambda_3$ obtained from  Eq.(\ref{eq:gcon1}) and Eq.(\ref{eq:gcon2})
is written as,
\bea
{\rm Max.}(-\frac{M_H}{v} \sqrt{\lambda_2}, -\frac{M_H}{v} \sqrt{\lambda_2}-2\frac{M_A^2-M_{H^+}^2}{v^2}) < \lambda_3.
\label{eq:l3}
\eea
One can also write $\lambda_3$ with the charged Higgs mass formulae, 
\bea
\lambda_3=\frac{2}{v^2}(M_{H^+}^2-m_{22}^2). 
\eea
Depending on the sign of $m_{22}^2$, the upper bound and the lower bound of 
$\lambda_3$ can be obtained for a given charged Higgs mass. Combining it with 
Eq.(\ref{eq:l3}), the constraints for positive $m_{22}^2$ case are,
\bea
{\rm Max.}(-\frac{M_H}{v} \sqrt{\lambda_2},-\frac{M_H}{v} \sqrt{\lambda_2}-2\frac{M_A^2-M_{H^+}^2}{v^2}) <\lambda_3< 
\frac{2 M_{H^+}^2}{v^2}, \quad (m_{22}^2>0).
\label{eq:cr1} 
\eea
When $m_{22}^2 \le 0$, in addition to the lower bound on $\lambda_3$,
the constraint on $\lambda_2$ in Eq.(\ref{eq:c12}) should be satisfied, 
\bea 
&&\frac{2 M_{H^+}^2}{v^2} \le \lambda_3,
\sqrt{\lambda_2} > (\lambda_3-2 \frac{M_{H^+}^2}{v^2}) \frac{v}{M_H}, 
\quad (m_{22}^2 <0).
\label{eq:cr2}
\eea

Now we study the quantum corrections numerically. 
We fix the standard
model like Higgs mass as $M_H=130$ (GeV).
There are still four parameters
to be fixed and they are $\lambda_2, \lambda_3, M_A$ and $M_{H^+}$.
Focusing on the Higgs mass spectrum of the extra Higgs, we study
the radiative corrections for  
the following scenarios for Higgs spectrum and the coupling constants.
\subsection{Case for $M_A=M_{H^+}$; degenerate charged Higgs and pseudoscalar Higgs and a relation for vanishing quantum correction  $\beta^{(1)}$}
We first study the corrections for degenerate charged Higgs and pseudoscalar
Higgs.  In this case, for a given degenerate mass, one can identify the 
values of
coupling constants $\lambda_2$ and $\lambda_3$ for which $\beta^{(1)}$ vanish. 
With $M_A=M_{H^+}$, the relation for coupling constants
which satisfies $\beta^{(1)}=0$ is,
\bea
\lambda_2&=&\frac{\lambda_3^2}{3 \lambda_1}
\left\{ 2+\frac{M_H^2}
{M_H^2-M_{H^+}^2} \left(1-\frac{M_H^2}{M_{H^+}^2} 
\frac{\log\frac{M_H^2}{\mu^2}-1}{\log\frac{M_{H^+}^2}{\mu^2}-1}\right) 
\right\} 
\nn  \\
&-&\frac{\lambda_3}{3}\left(\frac{M_{H^+}^2}{M_H^2-M_{H^+}^2}-
\frac{M_H^2}{M_H^2-M_{H^+}^2} \frac{M_H^2}{M_{H^+}^2} 
\frac{\log\frac{M_H^2}{\mu^2}-1}{\log\frac{M_{H^+}^2}{\mu^2}-1}\right).
\label{eq:rel}
\eea
The set of coupling constants $(\lambda_3, \lambda_2)$ which satisfy the relation Eq.(\ref{eq:rel}) are shown in table III.
We note that when $\lambda_2$ is as large as 10, $\lambda_3$ is at most 
about 3. If $\lambda_2$ is 1, $\lambda_3$ is lies in 
the range $0.55 \sim 0.7$.
\begin{table}
\begin{center}
\begin{tabular}{|c|c|c|c|} \hline
$\lambda_2$ & $\lambda_3$ ($M_{H^+}=100$) & $\lambda_3$ ($M_{H^+}=200$) & 
$\lambda_3$ ($M_{H^+}=500$)\\ \hline 
  0.14     & 0.19& 0.16  &0.18  \\ \hline
  0.28     & 0.28& 0.28  & 0.28 \\ \hline
  0.56     & 0.41& 0.47  & 0.42 \\ \hline
  1.0     & 0.55 & 0.69  & 0.59 \\ \hline
  10    & 1.8 & 2.8  & 2.0 \\ \hline
\end{tabular}
\caption{The coupling constants $(\lambda_3,\lambda_2)$ which satisfy the 
relation, Eq.(\ref{eq:rel}) for the three degenerate masses $M_{H^+}=M_A=100,200$ and $500$ (GeV).}
\end{center}
\end{table}
\begin{figure}[!htb]
\begin{center}
\includegraphics[width=15cm]{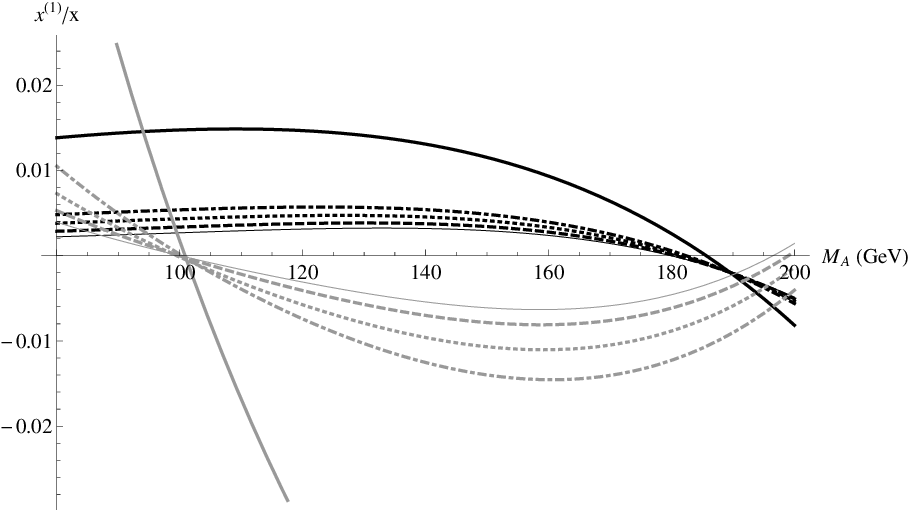} 
\caption{The quantum correction $\frac{\beta^{(1)}}{\beta}$ (gray lines) and 
$\frac{v^{(1)}}{v}$ (black lines) due to
the non-degeneracy of charged Higgs and pseudoscalar Higgs masses.
The pseudoscalar Higgs mass $M_A$ (GeV) dependence of the quantum 
corrections $\frac{x^{(1)}}{x}$ ($ x=\beta, v$) is shown
while the charged Higgs mass is fixed as $M_{H^+}=100$ (GeV). 
The set of parameters
$(\lambda_3, \lambda_2)$ are chosen so that the correction $\beta^{(1)}$ vanishes for the degenerate case; $M_{H^+}=M_A=100$ (GeV).
The values $(\lambda_3,\lambda_2)$ are taken from
Table (III) and they are   
$(0.19,0.14)$ (solid line), $(0.28,0.28)$ (dashed line), $(0.41,0.56)$ (dotted line), $(0.55,1)$ (dotdashed line), and $(1.8,10)$ (thick solid line).}
\end{center}
\label{fig1}
\end{figure}
\begin{figure}[!htb]
\begin{center}
\includegraphics[width=15cm]{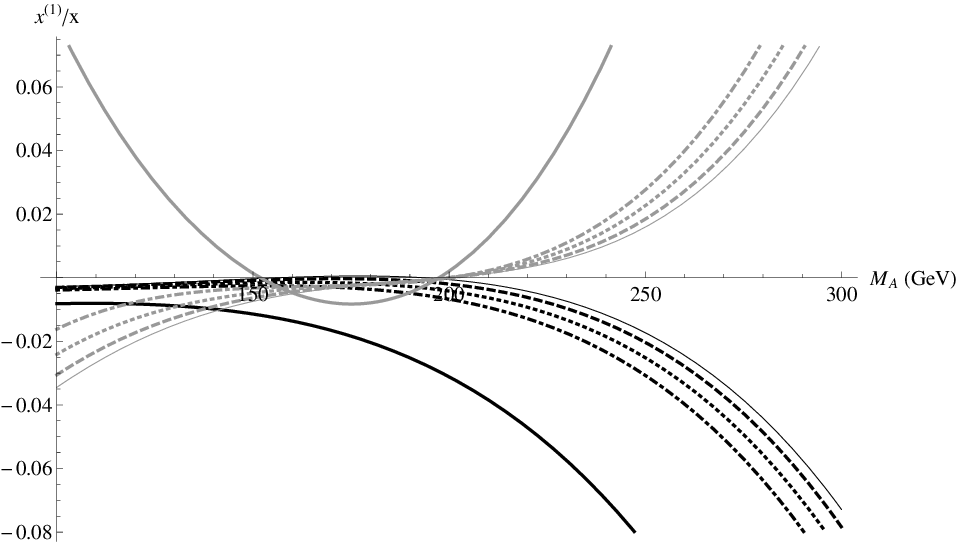} 
\caption{The quantum correction $\frac{\beta^{(1)}}{\beta}$ (gray lines) and 
$\frac{v^{(1)}}{v}$ (black lines) due to
the non-degeneracy of charged Higgs and pseudoscalar Higgs masses.
The pseudoscalar Higgs mass $M_A$ (GeV) dependence of the quantum 
corrections $\frac{x^{(1)}}{x}$ ($ x=\beta, v$) is shown
while charged Higgs mass is fixed as $M_{H^+}=200$ (GeV). The set of parameters
$(\lambda_3, \lambda_2)$ are chosen so that the correction $\beta^{(1)}$ vanishes for the degenerate case; $M_{H^+}=M_A=200$ (GeV).
The values $(\lambda_3,\lambda_2)$ are taken from
Table (III) and they are   
$(0.16,0.14)$ (solid line),
$(0.28,0.28)$ (dashed line)
, $(0.47,0.56)$ (dotted line), $(0.69,1)$ (dotdashed line), and $(2.8,10)$ 
(thick solid line).}
\end{center}
\label{fig2}
\end{figure}
\begin{figure}[!htb]
\begin{center}
\includegraphics[width=15cm]{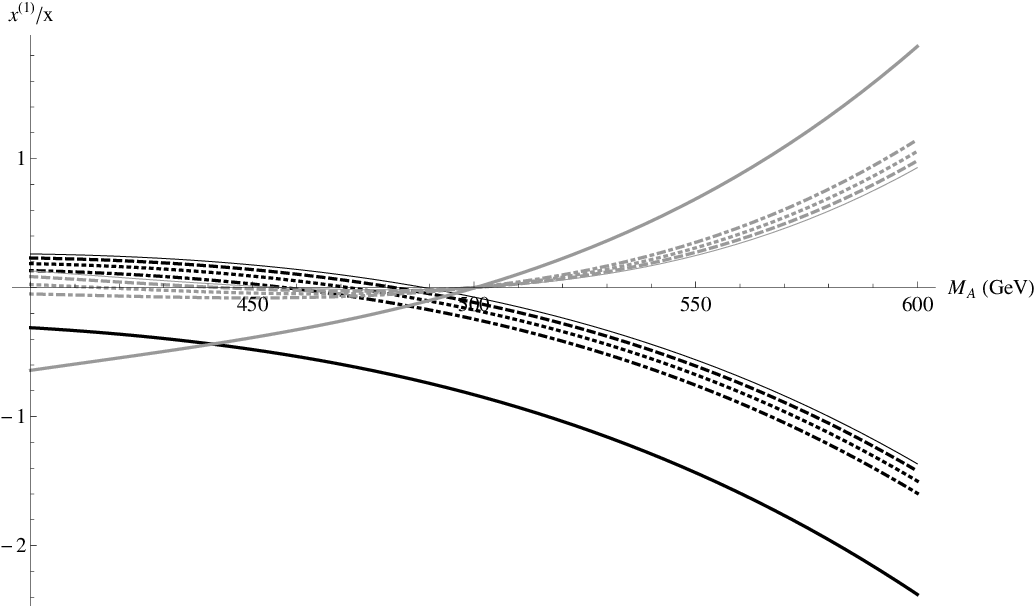} 
\caption{The quantum correction $\frac{\beta^{(1)}}{\beta}$ due to
the non-degeneracy of charged Higgs and pseudoscalar Higgs masses.
The pseudoscalar Higgs mass $M_A$(GeV) dependence of the quantum 
corrections $\frac{x^{(1)}}{x}$ ($x=\beta, v$) is shown
while charged Higgs mass is fixed as $M_{H^+}=500$ (GeV). The set of parameters
$(\lambda_3, \lambda_2)$ are chosen so that the correction $\beta^{(1)}$ vanishes for the degenerate case; $M_{H^+}=M_A=500$ (GeV).
The values $(\lambda_3,\lambda_2)$ are taken from
Table (III) and they are   
$(0.18,0.14)$(solid line), $(0.28,0.28)$ (dashed line), $(0.42,0.56)$ (dotted line), $(0.59,1)$ (dotdashed line), and $(2,10)$(thick solid line).}
\end{center}
\label{fig3}
\end{figure}
\subsection{Non-Degenerate case $M_A \ne M_{H^+}$ with the coupling constants 
satisfying Eq.(\ref{eq:rel})}
Next we lift the degeneracy by shifting the pseudoscalar Higgs mass
from the charged Higgs mass and study the effect on $\beta^{(1)}$ and 
$v^{(1)}$.
The non-degeneracy of the charged Higgs mass and the pseudoscalar Higgs mass
is constrained by $\rho$ parameter.
We change the pseudoscalar Higgs mass within the range $|M_A-M_{H^+}|<100$
(GeV) allowed from the electro-weak precision studies.
The coupling constants $(\lambda_3,\lambda_2)$ are chosen from
the sets of their values satisfying the relation Eq.(\ref{eq:rel}).
In Fig.1, we show $\frac{\beta^{(1)}}{\beta}$ as a function of
$M_A$ with charged Higgs mass $M_{H^+}=100$ (GeV).
When $M_A=100$ (GeV), the correction vanishes exactly.
As we increase $M_A$ from $100$ (GeV) (the mass of charged Higgs),
the correction becomes non-zero and is negative. The corrections are at most
about $1.3 \%$ when $\lambda_2 \sim 1$.
By increasing $M_A$ further, we meet the
point around at $M_A \simeq 200$ (GeV) corresponding to that 
the correction vanishes again.
In Fig.2, we study the correction $\beta^{(1)}$ with larger charged Higgs mass 
case, $M_{H^+}=200$(GeV). In contrast to the case for $M_{H^+}=100$ (GeV), 
by increasing $M_A$ from $200$ (GeV) where the correction vanishes, 
it increases and becomes positive. We also note that the correction
tend to be larger than the lighter charged Higgs mass case.
When $\lambda_2 \sim 1$, 
increasing  the pseudoscalar Higgs mass from 
$200$ (GeV) to $300$ (GeV), 
the correction is about
$10 \%$.
As the pseudoscalar Higgs mass decreases from $200$ (GeV) to $100$ (GeV), 
the correction becomes negative for $0<\lambda_2 \le 1$.  
With the larger value $\lambda_2=10$, we meet the point around at
$M_A \simeq 150$(GeV) where the correction vanishes again.
In Fig.3, we study 
the further larger 
charged Higgs mass case, 
i.e., $M_{H^+}=500$ (GeV). With $M_A \simeq 600$ (GeV),
the correction is positive and about 
$100\%$. The correction stays  small  
for $0<\lambda_2 \le 1$ when decreasing $M_A$
from $500$ (GeV) to $400$ (GeV).
\begin{figure}[!htb]
\begin{center}
\begin{tabular}{ccc}
\begin{minipage}{0.5\hsize}
\begin{center}
\includegraphics[height=8cm]{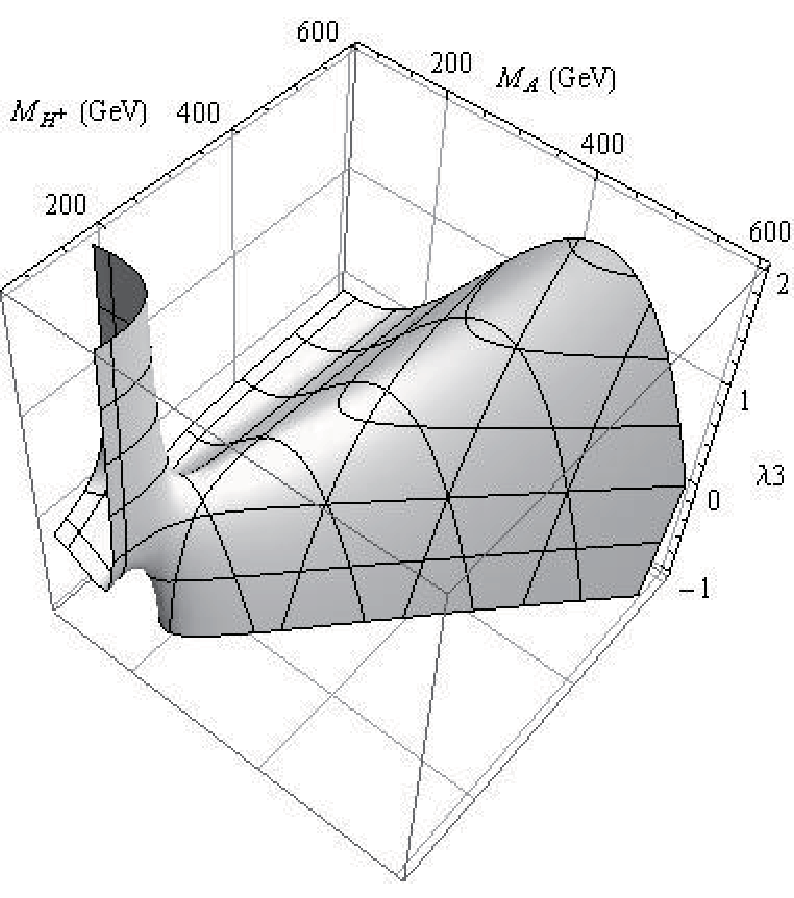}
\caption{The two dimensional surface for $v^{(1)}=0$.}
\label{fig4}
\end{center}
\end{minipage}&\quad \quad &
\begin{minipage}{0.5\hsize}
\begin{center}
\includegraphics[height=7cm]{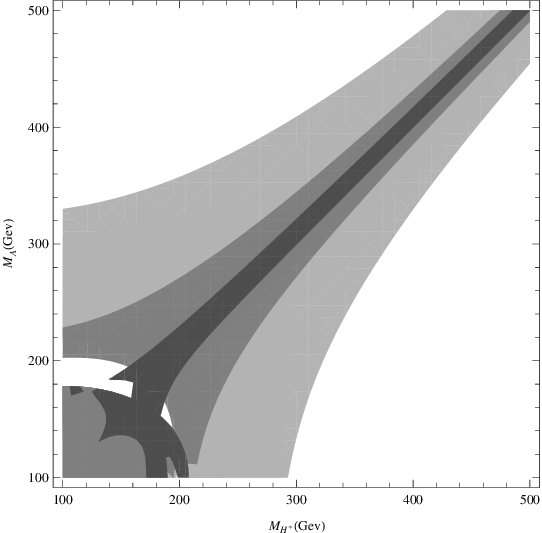}
\caption{The regions of ($M_{H^+},M_A$) which correspond to $\Bigl{(}\Bigr{|}\frac{v^{1}}{v}\Bigl{|},\Bigr{|}\frac{\beta^{(1)}}{\beta}\Bigl{|} \Bigr{)}=(0,0)$  (dark gray), $(0.01,0.01)$ (gray), and $(0.1,0.1)$ (light gray).}
\label{fig5}
\end{center}
\end{minipage}
\end{tabular}
\end{center}
\end{figure}
\subsection{The correction $\frac{v^{(1)}}{v}$}
In Figures 1,2, and 3, we also show the correction $\frac{v^{(1)}}{v}$ 
as functions of $M_A$. $v^{(1)}$ is independent on $\lambda_2$
and does not necessarily vanish at
the same points where $\beta^{(1)}$ vanishes.  With $\lambda_3 \ge 2$ and 
$M_{H^+} \ge 200$ (GeV), when the pseudoscalar Higgs mass 
is much larger than that of
charged Higgs mass, we find very large correction to $v$.
In Fig.4, we show that the two dimensional surface which corresponds 
to $v^{(1)}=0$. We find that the interior of the surface 
corresponds to the region of the positive correction; $v^{(1)}>0$ while the 
exterior region of 
the surface corresponds to the negative correction; $v^{(1)}<0$.\\

  In Fig.5, we have shown the regions of ($M_{H^+}, M_A$)
which correspond to 
that the corrections of $|v^{(1)}|$ and $|\beta^{(1)}|$ 
have the definite values ($0, 0.01, 0.1$).
The dark gray shaded area corresponds to the region where 
both $v^{(1)}$ and $\beta^{(1)}$ can vanish with taking account of the
conditions; Eq.(\ref{eq:gcon0}), Eq(\ref{eq:gcon1}) and Eq.(\ref{eq:gcon2}). 
We note that
for $M_{H^+}, M_A >200$ (GeV), the quantum corrections vanish
around the region where the charged Higgs degenerates with the pseudoscalar 
Higgs. 
When the corrections become larger, 
the larger mass splitting of the pseudoscalar Higgs
and charged Higgs is allowed.  However as the average mass of the
charged Higgs and pseudoscalar Higgs increases,
the allowed mass splitting
becomes smaller.
\section{discussion and conclusion}
In this paper, the Dirac neutrino mass model of Davidson and Logan is studied. 
In the model, one of the vacuum expectation values of two Higgs doublets is 
very small and it becomes the origin of the mass of neutrinos. 
The ratio of the small vacuum expectation value $v_2$ and that 
of the standard like 
Higgs $v_1$ is $\tan \beta=\frac{v_2}{v_1}$. Therefore $\tan \beta$ is very small and typically it is $O(10^{-9})$. The smallness of $\tan \beta$ is guaranteed  by the smallness of the soft breaking term of $U(1)^{\prime}$.

We have treated the soft breaking term as perturbation and calculated,
in particular, the vacuum expectation of Higgs in the
leading order of the perturbation precisely.
As summarized in Table I, only by including the soft breaking terms, one
can argue which of the local minima minimizes the potential and becomes the
global minimum.
 We have studied the global minimum of the tree level Higgs potential including the effect of the soft breaking term as perturbation.
 
Beyond the tree level, we study the quantum correction to the vacuum expectation values  and $\tan \beta$ in a quantitative way.
In one loop level, we confirmed
that tree level vacuum is stable, i.e., the order 
parameters which vanish at tree level 
do not have the vacuum expectation value
as quantum correction. 
In one loop level, we derived the exact formulae for the quantum correction to $\beta$ in the leading order of expansion of the soft breaking parameter $m_{12}^2$. We have confirmed not only that the loop correction to $\tan \beta$ is proportional to the soft breaking term but also found that 
the correction depends on the Higgs mass spectrum and  some combination of
the quartic coupling constants of the Higgs potential. 
Technically, we carried out the calculation of the one loop effective potential
by employing O(4) real representation for SU(2) Higgs doublets. 

Dependence of the corrections on the Higgs spectrum is studied numerically.
We first derive a relation of the coupling constants which corresponds to
the condition that the correction to $\beta$ vanishes for degenerate extra Higgs masses. 
Next, we study the effect of non-degeneracy of the charged Higgs and pseudoscalar 
Higgs on the correction. If the charged Higgs mass is as light as 
100 (GeV) $\sim$ 200 (GeV),
allowing the mass difference of charged Higgs and pseudoscalar Higgs is about $100$(GeV), the quantum corrections to both $\beta$ and $v$ are within 
a few $\%$ for $(\lambda_3,\lambda_2) \sim (0.5, 1)$. 
If the charged Higgs is heavy $M_{H^+}=500$ (GeV), a slight increase 
of the pseudoscalar
Higgs mass from the degenerate point leads to very large corrections to 
$\beta$ and $v$.

One can argue the size of the quantum corrections to the neutrino mass
of the model, because the ratio of the tree level neutrino mass and one loop 
correction can be written as,
\bea
\frac{m_{\nu}^{(1)}}{m_\nu}=
\frac{v^{(1)}}{v}+\frac{\beta^{(1)}}{\beta},
\label{eq:neutrino}
\eea
where we take account of the corrections only
due to Higgs vacuum expectation
values. The formulae Eq.(\ref{eq:neutrino}) implies that radiative correction
to neutrino mass is related to the Higgs mass spectrum. 
Therefore once Higgs mass spectrum is measured in LHC, 
one can compute the radiative correction to the mass of neutrinos
using the formulae Eq.(\ref{eq:neutrino}). 
\begin{acknowledgements}
We would like to thank D. Kimura for reading the manuscript
and Y. Kitadono with
discussion. The work of K. Tamai is supported by Hiroshima University 
research assistant fellowship. 
T. M. would like to thank the Aspen Center for Physics and NSF 
grant No. 1066293 where a part of the work completed.
The work of T. M. is supported by KAKENHI, 
Grant-in-Aid for Scientific Research(C) No.22440283
from JSPS, Japan. \\

{\bf Note added}\\

After submitting the paper, we are aware that the stability of the 
model studied in this paper 
was also discussed in \cite{Haba:2011fn}.  Compared to their analysis,
we derived the 1 loop effective potential taking into account  
all the interactions of Higgs sector while they consider
a part of the interactions and study the stability in a
qualitative way. Using the effective
potential, we carried out the quantitative analysis of 
the quantum corrections. 
\end{acknowledgements} 
\appendix
\section{Derivation of one-loop effective potential}
In this appendix, we give the details of the derivation of
the one-loop effective potential and the counterterm in Eq.(\ref{eq:V1loop}).
One can split $M^2(\phi)_{ij}$ in Eq.(\ref{eq:MTMphi})
into the diagonal part and the off diagonal part
as,
$
\delta M^2(\phi)_{ij}=M^2(\phi)_{ij}-M^2(\phi)_{ii} \delta_{ij}.
$
The divergent part of $V_{1 {\rm loop}}$ can be easily computed
by expanding it up to the second order of $\delta M^2$,
\bea
V_{1{\rm loop}}&=&V^{(1)}+V_c, \nn \\
V^{(1)}&=&\frac{\mu^{4-d}}{2} \int \frac{d^d k}{(2 \pi)^d i} 
{\rm Tr} {\rm Ln}\{ (D_{ii}^{0 -1}+M_{ii}^2(\phi)) 
\delta_{ij}+\delta M^2_{ij}
-\sigma_1 m_{12}^2 \}\nn \\
&=& \sum_{i=1}^8  \frac{\mu^{4-d}}{2} \int \frac{d^d k}{(2 \pi)^d i} 
{\ln} \{D_{ii}^{0 -1}+M_{ii}^2(\phi)\} \nn \\
&-& \sum_{i,j=1}^8 \frac{\mu^{4-d}}{4} \int \frac{d^d k}{(2 \pi)^d i} 
D_{ii} (\delta M^2-\sigma_1 m_{12}^2)_{ij} D_{jj} 
(\delta M^2-\sigma_1 m_{12}^2)_{ji}+ ...,
\eea
where,
\bea 
D_{ii}^{-1}&=&D_{ii}^{0 -1}+M_{ii}^2(\phi),\nn \\
&=&\Biggl{\{}
\begin{array}{c}
{M_{ii}^2+m_{11}^2-k^2} \quad (1 \le i \le 4), \\
{M_{ii}^2+m_{22}^2-k^2} \quad (5 \le i \le 8).
\end{array}
\eea
The diagonal parts of the propagators are given as, 
\bea
D_{ii}=\Biggl{\{}\begin{array}{c}
\frac{1}{M_{ii}^2+m_{11}^2-k^2} \quad (1 \le i \le 4), \\
\frac{1}{M_{ii}^2+m_{22}^2-k^2} \quad (5 \le i \le 8).
\end{array}
\eea
In the modified minimal subtraction scheme,
Feynman integration is carried out  with help of
the well known formulae of dimensional regularization,
\bea
\mu^{4-d}\frac{1}{2} \int \frac{d^dk}{(2 \pi)^d i} \log(m^2-k^2) 
=-\frac{1}{64 \pi^2 \bar{\epsilon}} m^4 
+\frac{m^4}{64 \pi^2} \left(\log\frac{m^2}{\mu^2}-\frac{3}{2}
\right),
\label{eq:integ}
\eea
and,
\bea
\mu^{4-d}\int \frac{d^dk}{(2 \pi)^d i} \frac{1}{(m_{i}^2-k^2)
(m_{j}^2-k^2)}\Biggr{|}_{\rm div.}
=\frac{1}{16 \pi^2} \frac{1}{\overline{\epsilon}},
\eea
with $\frac{1}{\overline{\epsilon}}=\frac{1}{\epsilon}-\log 4\pi$ and 
$\epsilon=2-\frac{d}{2}$.
The divergent part of $V^{(1)}$ is,
\bea
V^{(1)}_{\rm div.}=&-&\frac{1}{64 \pi^2 \bar{\epsilon}} 
\{ \sum_{i=1}^4  (M^2_{ii}+m_{11}^2)^2+ \sum_{i=5}^8
(M^2_{ii}+m_{22}^2)^2 \} \nn \\
&-&\frac{1}{64 \pi^2 \bar{\epsilon}} \sum_{i \ne j=1}^8
(\delta M^2-m_{12}^2 \sigma_1)_{ij} (\delta M^2-m_{12}^2 \sigma_1)
_{ji},  \nn \\
=&-&\frac{1}{32 \pi^2 \bar{\epsilon}} 
\left(m_{11}^2 \sum_{i=1}^4 M^2_{ii}(\phi) +
m_{22}^2 \sum_{i=5}^8 M^2_{ii}(\phi) 
+2(m_{11}^4 + m_{22}^4)\right) \nn \\
&-&\frac{1}{64 \pi^2 \bar{\epsilon}} {\rm Tr}\Bigl{[}
(M^2(\phi)-m_{12}^2 \sigma_1)(M^2(\phi)-m_{12}^2 \sigma_1)
\Bigr{]}, \nn \\
=&-&\frac{1}{64 \pi^2 \bar{\epsilon}} 
{\rm Tr}[M_T^4].
\label{eq:div}
\eea

The trace of Eq.(\ref{eq:div}) is calculated in
Eq.(\ref{eq:decom0}) and Eq.(\ref{eq:MMMM}) of appendix B
and the result is, 
\bea
V^{(1)}_{\rm div.}
=&-&\frac{1}{32\pi ^2 \bar {\epsilon }} 
\Biggl{[}m_{11}^2 \{ 6\lambda _1 (\Phi _1^\dagger \Phi _1 )
+2(2\lambda _3+\lambda _4)(\Phi _2^\dagger \Phi _2) \}
\nn \\ 
&+&m_{22}^2 \{ 2(2\lambda _3+\lambda _4)(\Phi _1^\dagger \Phi _1) + 6\lambda _2 (\Phi _2^\dagger \Phi _2) \} \Biggr{]} \nn \\
&+& \frac{2m_{12}^2}{64\pi ^2\bar {\epsilon} }  
\Biggl{[}( 2\lambda _3+4\lambda _4) (\Phi _1^\dagger \Phi _2 + \Phi _2^\dagger \Phi _1)\Biggl{]} \nn \\
&-&\frac{8m_{12}^4+4 (m_{11}^4+m_{22}^4)}{64\pi ^2\bar {\epsilon }} \nn \\
&-& \frac{1}{64\pi ^2\bar {\epsilon }}  \Biggl{[} 
(12\lambda _1^2 + 4 \lambda _3\lambda _4 + 4\lambda _3^2 + 2\lambda _4^2)(\Phi _1^\dagger \Phi _1)^2 \nn \\
&+& (12\lambda _2^2 + 4 \lambda _3\lambda _4 + 4\lambda _3^2 + 2\lambda _4^2)(\Phi _2^\dagger \Phi _2)^2 \nn \\ 
&+& (12\lambda _1 \lambda _3 + 4\lambda _1 \lambda _4 + 8\lambda _3^2 + 4\lambda _4^2 + 12 \lambda _2 \lambda _3 
+ 4\lambda _2 \lambda _4)(\Phi _1^\dagger \Phi _1)(\Phi _2^\dagger \Phi _2) \nn \\
&+& (4\lambda _1\lambda _4+16\lambda _3\lambda _4+ 8\lambda _4^2 +4\lambda _2\lambda _4)\mid \Phi _1^\dagger \Phi _2 \mid^2 \Biggr{]}.
\label{eq:veffdiv}
\eea
Now the counterterms for the one loop effective potential
are simply given by changing the sign of the 
divergent part of Eq.(\ref{eq:veffdiv}),
\bea
V_c&=&-V^{(1)}_{\rm div.}\nn \\
   &=&\frac{1}{64 \pi^2 \bar{\epsilon}} 
{\rm Tr}[M_T^4].
\label{eq:counterterms}
\eea
Using Eq.(\ref{eq:counterterms}) and Eq.(\ref{eq:integ}), one can  
derive the finite part of the 1 loop effective potential given in Eq.(\ref{eq:V1loop}).
\section{Derivation of Eq.(\ref{eq:veffdiv})}
In this section, we present the derivation of Eq.(\ref{eq:veffdiv}).
We start with the quartic interaction terms of the Higgs potential,
\bea
V^{(4)}&=&\frac{\lambda_1}{8} (\sum_{i=1}^4 {\phi_i}^2)^2
+ \frac{\lambda_2}{8} (\sum_{i=5}^8 {\phi_i}^2)^2 
+ \frac{\lambda_3}{4} (\sum_{i=1}^4 {\phi_i}^2)   
(\sum_{j=5}^8 {\phi_j}^2) \nn \\
&+& \frac{\lambda_4}{4} \left(
(\phi_1 \phi_5 + \phi_2 \phi_6 + \phi_3 \phi_7 + \phi_4 \phi_8)^2
+(\phi_1 \phi_6+ \phi_3 \phi_8 -\phi_2 \phi_5 -\phi_4 \phi_7)^2 \right).
\eea
By taking the derivatives of $V^{(4)}$,
one can obtain the mass squared matrix
$M^2(\phi)$.  One first computes the first derivative
of $V^{(4)}$ with respect to $\phi_i$,
\bea
\frac{\partial V^{(4)}}{\partial \phi_i}=
\begin{cases}
\frac{\lambda_1}{8} 2(\sum_{j=1}^4 \phi_j^2) 2 \phi_i 
+ \frac{\lambda_3}{2} \phi_i \sum_{j=5}^8 \phi_j^2
+\frac{\lambda_4}{2}\{ (\phi_1 \phi_5 + \phi_2 \phi_6 + \phi_3 \phi_7
+ \phi_4 \phi_8) \phi_{i+4}  \nn \\
+( \phi_1 \phi_6 + \phi_3 \phi_8 
-\phi_2 \phi_5 -\phi_4 \phi_7) (\delta_{1i} \phi_6 -\delta_{2i} \phi_5 
+\delta_{3i} \phi_8 -\delta_{4i} \phi_7) \},
\ (1 \le i \le 4) \\      
\frac{\lambda_2}{8} 2(\sum_{j=5}^8 \phi_j^2) 2 \phi_i 
+ \frac{\lambda_3}{2} \phi_i \sum_{j=1}^4 \phi_j^2 
+ \frac{\lambda_4}{2} \{ (\phi_1 \phi_5 + \phi_2 \phi_6 + \phi_3 \phi_7
+ \phi_4 \phi_8) \phi_{i-4}   \nn \\
+ ( \phi_1 \phi_6 + \phi_3 \phi_8 -\phi_2 \phi_5 -\phi_4 \phi_7) (-\delta_{5i} \phi_2 +\delta_{6i} \phi_1 
-\delta_{7i} \phi_4 +\delta_{8i} \phi_3) \}.
\ (5 \le i \le 8).
\end{cases} 
\nn \\
\eea
The second derivatives are given as,
\bea
&& \frac{\partial^2 V^{(4)}}{\partial \phi_i \partial \phi_j}=
\nn \\
&& \begin{cases}
\frac{\lambda_1}{2} \left( \delta_{ij} \sum_{k=1}^4 \phi_k^2+ 2 \phi_j \phi_i
\right) + \frac{\lambda_3}{2} \delta_{ij} 
(\sum_{k=5}^8 \phi_k^2)
+ \frac{\lambda_4}{2} 
\left\{ \phi_{j+4} \phi_{i+4} + \right. \nn \\
\left. (\delta_{1j} \phi_6
-\delta_{2j} \phi_5 + \delta_{3j} \phi_8 -\delta_{4j} \phi_7)
(\delta_{1i} \phi_6 -\delta_{2i} \phi_5 + \delta_{3i} \phi_8
-\delta_{4i} \phi_7) \right\}, \ (1 \le i,j \le 4),\\
\lambda _3 \phi _i \phi _j 
+  \frac{\lambda _4}{2} \{ \phi _{i+4} \phi _{j-4} +\sum_{k=1}^4 \delta _{i+4,j}\phi _k \phi _{k+4}+
(-\delta _{5j} \phi _2 +\delta _{6j} \phi _1 -\delta _{7j} \phi _4 +\delta _{8j} \phi _3 ) \nn \\
(\delta _{1i} \phi _6 -\delta _{2i} \phi _5 +\delta _{3i} \phi _8 -\delta _{4i} \phi _7 ) +
(\phi _1 \phi _6 + \phi _3 \phi _8 - \phi _2 \phi _5 - \phi _4 \phi _7)
\nn \\
(\delta _{1i}\delta _{6j} +\delta _{3i}\delta _{8j}-\delta _{2i}\delta _{5j}-\delta _{4i}\delta _{7j}) \},
\quad (1\le i \le 4,  5 \le j \le 8),\nn \\
\lambda _3 \phi _i \phi _j + \frac{\lambda _4}{2} \{ \phi _{i-4} \phi _{j+4} +\sum_{k=1}^4 \delta _{i-4,j}\phi _k \phi _{k+4}
+ (\delta _{1j} \phi _6 -\delta _{2j} \phi _5 +\delta _{3j} \phi _8 -\delta _{4j} \phi _7 ) \times \nn \\
(-\delta _{5i} \phi _2 +\delta _{6i} \phi _1 -\delta _{7i} \phi _4 +\delta _{8i} \phi _3 ) +
(\phi _1 \phi _6 + \phi _3 \phi _8 - \phi _2 \phi _5 - \phi _4 \phi _7) \times
\nn \\
(\delta _{1i}\delta _{6j} +\delta _{3i}\delta _{8j}-\delta _{2i}\delta _{5j}-\delta _{4i}\delta _{7j}) \}, \quad (5 \le i  \le 8, 1 \le j \le 4),\nn \\
\frac{\lambda_2}{2} \left( \delta_{ij} 
\sum_{k=5}^8 \phi_k^2+ 2 \phi_j \phi_i
\right) + \frac{\lambda_3}{2} \delta_{ij} 
(\sum_{k=1}^4 \phi_k^2)
+ \frac{\lambda_4}{2} \{ \phi_{j-4} \phi_{i-4}+\nn \\
(-\delta_{5j} \phi_2 + \delta_{6j} \phi_1 - \delta_{7j} \phi_4 +\delta_{8j} \phi_3)
(-\delta_{5i} \phi_2 +\delta_{6i} \phi_1 - \delta_{7i} \phi_4 + \delta_{8i} \phi_3) \},
\ (5 \le i,j \le 8).
\end{cases} \nn \\
\label{eq:second}
\eea
With Eq.(\ref{eq:second}), 
the diagonal sums of $M^2$ are given as,
\bea
\sum_{i=1}^4M^2_{ii}&=&
3 \lambda _1 \sum_{i=1}^4 \phi _i^2 + 2\lambda _3 \sum_{i=5}^8 \phi _i^2 +\lambda _4 \sum_{i=5}^8 \phi _i^2
= 6\lambda _1 \Phi _1^\dagger \Phi _1 +(4\lambda _3+2\lambda _4)\Phi _2^\dagger \Phi _2, (1\leq i \leq  4),\nn \\
\sum_{i=5}^8M^2_{ii}&=&
3 \lambda _2 \sum_{i=5}^8 \phi _i^2 + 2\lambda _3 \sum_{i=1}^4 \phi _i^2 +\lambda _4 \sum_{i=1}^4 \phi _i^2 
= 6\lambda _2 \Phi _2^\dagger \Phi _2 +(4\lambda _3+2\lambda _4)\Phi _1^\dagger \Phi _1, (5\leq i \leq 8).\nn \\
\label{eq:B4}
\eea
The counterterm in Eq.(\ref{eq:counterterms}) includes the 
following contribution, 
\bea
{\rm Tr}[(M^2(\phi )-m_{12}^2 \sigma_1) (M^2(\phi)-m_{12}^2 \sigma_1)]
={\rm Tr}[M^2(\phi ) M^2(\phi ) -2m_{12}^2 \sigma_1 M^2] +8 m_{12}^4.
\label{eq:ct}
\eea 
The second term of Eq.(\ref{eq:ct}) is proportional to, 
\bea
{\rm Tr}[ m_{12}^2 \sigma_1 M^2] &=& (2\lambda _3+4\lambda _4) (\phi _1 \phi _5 + \phi _2 \phi _6 + \phi _3 \phi _7 + \phi _4 \phi _8 ) m_{12}^2 \nn \\
&=& (2\lambda _3+4\lambda _4) (\Phi _1^\dagger \Phi _2 + \Phi _2^\dagger  \Phi _1) m_{12}^2.
\label{eq:decom0}
\eea
The first term of Eq.(\ref{eq:ct}) can be decomposed as, 
\bea
{\rm Tr}[M^2(\phi)M^2(\phi)]&=&\sum_{i,j=1}^4
M^2(\phi)_{ij}M^2(\phi)_{ji}+
2 \sum_{i=1}^4 \sum_{j=5}^8 M^2(\phi)_{ij}M^2(\phi)_{ji}
\nn \\ 
&+&\sum_{i,j=5}^8 M^2(\phi)_{ij}M^2(\phi)_{ji}.
\label{eq:decom}
\eea
Each term of Eq.(\ref{eq:decom}) is given as,
\bea
&& \sum_{i,j=1}^4M^2(\phi )_{ij} M^2(\phi )_{ji} = 3 \lambda _1^2 \left( \sum_{i=1}^4 \phi _i^2 \right)^2 
+ 3 \lambda _1\lambda _3  \sum _{i=1}^4 \phi _i^2 \sum _{j=5}^8 
\phi _j^2 
+ \lambda _1\lambda _4 
\left\{  \sum_{i=5}^8 \phi _i^2 \sum _{j=1}^4 \phi _j^2 
\right. \nn \\
&& \ \left.+
(\phi _1 \phi _5 + \phi _2 \phi _6 + \phi _3 \phi _7 + \phi _4 \phi _8 )^2
+ ( \phi_1 \phi_6 + \phi_3 \phi_8 -\phi_2 \phi_5 -\phi_4 \phi_7)^2 
\right\} \nn \\
&& \ + 
\lambda _3\lambda _4 \left( \sum _{i=5}^8 \phi _i^2 \right) ^2
+ 
\lambda _3^2 \left( \sum _{i=5}^8 \phi _i^2 \right) ^2 + \frac{\lambda _4^2}{2} \left( \sum _{i=5}^8 \phi _i^2 \right) ^2 \nn \\
&& \ = 12 \lambda _1^2 ( \Phi _1^\dagger \Phi _1 )^2 
+ (12\lambda _1 \lambda _3 +4 \lambda _1\lambda _4)(\Phi _1^\dagger \Phi _1)(\Phi _2^\dagger \Phi _2)\nn \\ 
&& \ + 4\lambda _1 \lambda _4\mid \Phi _1^\dagger \Phi _2 \mid^2
+ (4\lambda _3 \lambda _4 +4\lambda _3^2 +2\lambda _4^2)(\Phi _2^\dagger \Phi _2)^2, \label{eq:decom3-1}\\
&&\sum_{i=1}^4 \sum_{j=5}^8 M^2(\phi )_{ij} M^2(\phi )_{ji}= \lambda _3^2 \sum_{i=5}^8 \phi _i^2 \sum _{j=1}^4 \phi _j^2 
+2\lambda _3\lambda _4 \left\{ \sum_{i=1}^4 \phi _i\phi _{i+4} \sum_{j=1}^4 \phi _j \phi _{j+4} \right. \nn \\
&& \ + \left. (\phi _1\phi _6 - \phi _2\phi _5 + \phi _3 \phi _8 - \phi _4\phi _7)^2 \right\} \nn \\
&& \ + \frac{\lambda _4^2}{2}\left\{ \sum_{i=1}^4 \phi _i^2 \sum_{j=5}^8\phi _j^2+2\left( \sum_{i=1}^4\phi _i \phi _{i+4} \right)^2
+ 2(\phi _1\phi _6-\phi _2\phi _5 + \phi _3\phi _8-\phi _4\phi _7)^2 \right\} \nn \\
&& \ = \left( 4\lambda _3^2 +2\lambda _4^2 \right) (\Phi _1^\dagger \Phi _1)(\Phi _2^\dagger \Phi _2) 
+ (8 \lambda _3\lambda _4 + 4\lambda _4^2 ) \mid \Phi _1^\dagger \Phi _2 \mid ^2, \label{eq:decom3-2}\\
&& \sum_{i,j=5}^8 M^2(\phi )_{ij} M^2(\phi )_{ji} = 3 \lambda _2^2 \left( \sum_{i=5}^8 \phi _i^2 \right)^2 
+ 3 \lambda _2\lambda _3  
\sum _{i=5}^8 \phi _i^2 \sum _{j=1}^4 \phi _j^2 
+ \lambda _2\lambda _4 \left\{  \sum_{i=1}^4 \phi _i^2 
\sum _{j=5}^8 \phi _j^2 + \right.\nn \\ 
&& \ \left. 
(\phi _1 \phi _5 + \phi _2 \phi _6 + \phi _3 \phi _7 + \phi _4 \phi _8 )^2
+( \phi_1 \phi_6 + \phi_3 \phi_8 -\phi_2 \phi_5 -\phi_4 \phi_7)^2 
\right\} \nn \\
&& \ + \lambda _3\lambda _4 
\left( \sum _{i=1}^4 \phi _i^2 \right) ^2 
+ \lambda _3^2 \left( \sum _{i=1}^4 \phi _i^2 \right) ^2 
+ \frac{\lambda _4^2}{2} \left( \sum _{i=1}^4 \phi _i^2 \right) ^2 \nn \\
&& \ = 12 \lambda _2^2 ( \Phi _2^\dagger \Phi _2 )^2 
+ (12\lambda _2 \lambda _3 +4 \lambda _2\lambda _4)(\Phi _1^\dagger \Phi _1)(\Phi _2^\dagger \Phi _2) \nn \\ 
&& \ +4\lambda _2 \lambda _4\mid \Phi _1^\dagger \Phi _2 \mid^2 + 
(4\lambda _3 \lambda _4 +4\lambda _3^2 +2\lambda _4^2)(\Phi _1^\dagger \Phi _1)^2.
\label{eq:decom3-3}
\eea
From Eq.(\ref{eq:decom3-1}), Eq.(\ref{eq:decom3-2}),and
Eq.(\ref{eq:decom3-3}), one obtains,
\bea
{\rm Tr}[M^2(\phi ) M^2(\phi )] &=& (12\lambda _1^2 + 4 \lambda _3\lambda _4 + 4\lambda _3^2 + 2\lambda _4^2)(\Phi _1^\dagger \Phi _1)^2 \nn \\
&+& (12\lambda _2^2 + 4 \lambda _3\lambda _4 + 4\lambda _3^2 + 2\lambda _4^2)(\Phi_2^\dagger \Phi _2)^2 \nn \\ 
&+& (12\lambda _1 \lambda _3 + 4\lambda _1 \lambda _4 + 8\lambda _3^2 + 4\lambda _4^2 + 12 \lambda _2 \lambda _3 
+ 
4\lambda _2 \lambda _4)(\Phi _1^\dagger \Phi _1)(\Phi _2^\dagger \Phi _2) \nn \\
&+& (4\lambda _1\lambda _4+16\lambda _3\lambda _4+ 8\lambda _4^2 +4\lambda _2\lambda _4)\mid \Phi _1^\dagger \Phi _2 \mid^2.  
\label{eq:MMMM}
\eea
Using  Eq.(\ref{eq:B4}), Eq.(\ref{eq:ct}), Eq.(\ref{eq:decom0}), and Eq.(\ref{eq:MMMM}),
one can derive Eq.(\ref{eq:veffdiv}).
\section{$[O^T \frac{\partial M^2}{\partial \varphi_I} O]_{jj}$ 
and $L_{IJ}$}
In this appendix, we show $
[O^T \frac{\partial M^2}{\partial \varphi_I} O]_{jj}$ and $L_{IJ}$
which are needed to calculate one loop corrections to the order 
parameters $\varphi_I^{(1)}$ in Eq.(\ref{eq:ordersimp}).
$
[O^T \frac{\partial M^2}{\partial \varphi_I} O]_{jj}$ ($I=1,2,3,4$) are given as,
\bea
&&[O^T \frac{\partial M^2}{\partial \alpha} O]_{jj}=0,\quad
[O^T \frac{\partial M^2}{\partial \theta^\prime} O]_{jj}=0. 
\label{eq:althp}
\eea
\bea
&&[O^T \frac{\partial M^2}{\partial v} O]_{jj}=2 v [O^T \frac{\partial M^2}{\partial v^2} O]_{jj}
= \frac{v}{4} \times \nn \\
&&\begin{pmatrix}
\frac{1}{2} \left(\lambda _1+\lambda _2+6 \lambda _3-2 \lambda _4-\cos (4 \beta ) \left(\lambda _1+\lambda _2-2 \left(\lambda
   _3+\lambda _4\right)\right)\right) \\
\frac{1}{2} \left(\lambda _1+\lambda _2+
6 \lambda _3-2 \lambda _4-\cos (4 \beta ) \left(\lambda _1+\lambda _2-2 \left(\lambda
   _3+\lambda _4\right)\right)\right)  \\
\frac{1}{2} \left(\lambda _1+\lambda _2+6 \lambda _3+6
\lambda _4-\cos (4 \beta ) \left(\lambda _1+\lambda _2-2 \left(\lambda
   _3+\lambda _4\right)\right)\right)\\ 
12\{\lambda_2 \cos^2\gamma  \sin^2 \beta+
\cos^2 \beta \sin^2 \gamma  \lambda _1\}+(3 \cos 2
   (\beta -\gamma )-\cos 2 (\beta +\gamma )+2) 
\left(\lambda _3+\lambda _4\right) \\
12\{\lambda_1 \cos ^2 \beta   \cos^2 \gamma+ \sin^2 \beta 
\sin^2 \gamma \lambda _2\}+(-3 \cos 2
   (\beta -\gamma )+\cos 2 (\beta +\gamma )+2) 
\left(\lambda _3+\lambda _4\right)
\end{pmatrix}, \nn \\
\label{eq:v}
\eea
and,
\bea
&&[O^T \frac{\partial M^2}{\partial \beta} O]_{jj}=v^2 
\frac{\sin 2 \beta}{2}\nn \\
&&
\begin{pmatrix}
\lambda _2 \cos ^2(\beta )- \sin ^2(\beta ) \lambda _1-
 \cos (2 \beta )
   \left(\lambda _3+\lambda _4\right)\\
 \lambda _2 \cos ^2(\beta )- \sin ^2(\beta ) \lambda _1-
 \cos (2 \beta )
   \left(\lambda _3+\lambda _4\right)\\
 \lambda _2 \cos ^2(\beta )- \sin ^2(\beta ) \lambda _1-
 \cos (2 \beta )
   \left(\lambda _3+\lambda _4\right)\\
 3  \lambda _2 \cos ^2(\gamma )-3  \sin^2 \gamma
\lambda_1+\frac{1}{2 \sin 2 \beta} 
(\sin (2 (\beta +\gamma ))-3 \sin (2 (\beta -\gamma ))) \left(\lambda _3+\lambda _4\right)\\
-3  \lambda _1 \cos ^2(\gamma )+3 \sin ^2(\gamma ) 
\lambda _2 - \frac{1}{ 2 \sin 2 \beta}
(\sin (2 (\beta
   +\gamma ))-3 \sin (2 (\beta -\gamma ))) 
\left(\lambda _3+\lambda _4\right)
\end{pmatrix},\nn \\
\label{eq:bet}
\eea
Next we show $L_{IJ}$ in Eq.(\ref{eq:LIJ}). Note that $L_{IJ}$ is symmetric
$L_{IJ}=L_{JI}$ and 
its non-zero elements are,
\bea
L_{11}&=&
\cos^2\beta m_{11}^2+\sin^2\beta m_{22}^2
-2 \cos (\beta ) \sin (\beta)m_{12}^2 \nn \\
&+& \frac{1}{2} 
\Biggl{[}3 v^2 \{ \lambda _1 \cos ^4(\beta ) 
+\sin ^2(\beta ) 
\left(2 \left(\lambda _3+\lambda _4\right) \cos ^2(\beta )+
\sin ^2(\beta ) \lambda _2\right)\} \Biggr{]}, \nn \\
L_{22}&=&v^2
\{-\frac{\cos 4 \beta}{4} \left(\lambda _1+\lambda _2-2 (\lambda _3+\lambda _4)
\right) v^2+ \frac{\cos 2\beta}{4} (\lambda _2-\lambda _1) v^2
\nn \\ 
&+&
2  m_{12}^2 \sin 2\beta  - \cos 2\beta (m_{11}^2-m_{22}^2)\},
      \nn \\
L_{12}&=&L_{21}=v \{-\frac{\sin 4\beta}{4} 
\left(\lambda _1+\lambda _2-2 \left(\lambda _3+\lambda _4\right)\right) v^2
+\frac{1}{2} \sin 2\beta \left(\lambda _2-\lambda _1\right) v^2,
\nn \\
&-&2  m_{12}^2\cos 2\beta
-\sin 2\beta (m_{11}^2-m_{22}^2) \}  \nn \\
L_{33}&=&-\frac{1}{8} v^2 \sin (2 \beta ) \left(v^2 \sin (2 \beta ) \lambda _4-4 m_{12}^2\right),\nn \\
L_{44}&=&v^2 \cos (\beta ) \sin (\beta ) m_{12}^2.
\label{eq:elementofL}
\eea
\section{Orthogonal matrix $O$ in Eq.(\ref{eq:Higgsmass})}
Here we show the orthogonal matrix $O$ in Eq.(\ref{eq:Higgsmass}).
\bea
O=\begin{pmatrix}
0 & -\sin \beta& 0 & 0 & 0& 0 & \cos \beta& 0 \\
-\sin \beta & 0 & 0 & 0 & 0 &\cos \beta &0 & 0\\
0&0 & 0 & \sin \gamma & \cos \gamma & 0 &0 &0 \\
0&0& -\sin \beta & 0 & 0& 0 & 0& \cos \beta \\
0& \cos \beta & 0 &0  & 0& 0 & \sin \beta &0\\
\cos \beta &0 & 0 & 0 & 0& \sin \beta  &0 & 0 \\
0 &0 & 0 & \cos \gamma & -\sin \gamma & 0 & 0& 0\\
0 &0 & \cos \beta  & 0 & 0& 0 &0 & \sin \beta
\end{pmatrix}.
\eea
\bibliographystyle{unsrt}

\end{document}